\documentclass[prl,reprint,preprintnumbers]{revtex4-1}
\usepackage{aj-definitions}

\usepackage[bbgreekl]{mathbbol} 
\usepackage{slashed} 

\newcommand{\lettersection}[1]{\section{#1}}

\begin{document}

\preprint{DCPT-16/41}

\title{A theory of non-Abelian superfluid dynamics} \author{Akash Jain}
\email[]{akash.jain@durham.ac.uk, ajainphysics@gmail.com} \affiliation{Department of Mathematical
  Sciences \& Centre for Particle Theory, Durham University, Durham - DH1 3LE, UK.}


\begin{abstract}
  We write down a theory for non-Abelian superfluids with a partially broken (semisimple) Lie group.
  We adapt the offshell formalism of hydrodynamics to superfluids and use it to comment on the
  superfluid transport compatible with the second law of thermodynamics. We find that the second law
  can be also used to derive the Josephson equation, which governs dynamics of the Goldstone
  modes. In the course of our analysis, we derive an alternate and mutually distinct parametrization
  of the recently proposed classification of hydrodynamic transport and generalize it to
  superfluids.
\end{abstract}

\maketitle

\countstart


Hydrodynamics is the study of universal low energy fluctuations of a quantum system near its ground
state. Any quantum system in this regime, called a \emph{fluid}, can be characterized by a set of
transport coefficients such as pressure, viscosity and conductivity.
When a part of the global symmetry of the microscopic theory is spontaneously broken in the ground
state, low energy fluctuations can also contain massless Goldstone modes \cite{Goldstone:1961eq}
corresponding to the broken symmetry. Therefore the associated fluid, commonly known as a
\emph{superfluid} \cite{landau1959fluid,putterman1974superfluid,Son:2000ht}, contains many new
transport coefficients in its spectrum. Superfluidity with a broken $\rmU(1)$ was first observed in
liquid $^4$He \cite{1938Natur.141...74K,1938Natur.142..643A}, which since then has been well
explored in the literature, at least up to the first order in derivatives (see
e.g. \cite{Bhattacharya:2011tra,Neiman:2011mj}). In recent years, non-Abelian superfluids have also
started to attract some attention (see \cite{Hoyos:2014nua} and references therein) in relation to
the $p$-wave superfluidity observed in liquid $^3$He
\cite{PhysRevLett.28.885,PhysRevLett.29.920}. On a different front, entire transport of an ordinary
fluid compatible with the second law of thermodynamics has been classified
\cite{Haehl:2014zda,Haehl:2015pja}, and a good amount of progress is being made towards writing down
a Wilsonian effective action describing the entire ordinary hydrodynamics
\cite{Haehl:2015pja,Haehl:2015uoc,Crossley:2015tka,deBoer:2015ija}.

The goal of this note is to set up a theory for superfluids with an arbitrarily broken internal
symmetry, and explore the constraints imposed upon it by the second law of thermodynamics. In
particular, we will show how the Josephson equation, which governs dynamics of the Goldstone modes,
naturally emerges in our formalism as a consequence of the second law. While addressing these
questions, we will propose a natural and mutually distinct classification of the entire (super)fluid
transport, which in the ordinary fluid limit gives a refined parametrization of the classification
mentioned above \cite{Haehl:2014zda,Haehl:2015pja}.


\lettersection{Spontaneous symmetry breaking}Let us start with a quick recap of the spontaneous
symmetry breaking; details can be found in \S 19 of \cite{Weinberg:1996kr}. Consider a microscopic
theory invariant under spacetime translations and action of a spacetime invariant semisimple Lie
group $G$ (with Lie algebra $i\fg$). Let $\p$ be a field in the theory transforming under some
unitary representation $\cD(G)$ of $G$, i.e. under a $g\in G$ transformation $\p \ra \cD(g)\p$. $\p$
is said to spontaneously break the symmetry from $G$ to its Lie subgroup $H \subset G$ (with Lie
subalgebra $i\fh\subset i\fg$), if its ground state expectation value $\<\p\>$ is only invariant
under $H$, i.e.  $\cD(h)\<\p\> = \<\p\>$ if and only if $h\in H$. $\cD(g)\<\p\>$ with $g\notin H$
are ``other'' ground states system could have spontaneously chosen from. Around $\<\p\>$, the field
$\p$ can be expressed as group transformation of a reference field $\tilde\p$, i.e.
$\p = \cD(\g)\tilde\p$, defined by,
\begin{equation}\label{tildepdefn}
  \tilde\p^\dagger \cD(g) \<\p\> = \tilde\p^\dagger \<\p\>, \qquad \forall \ g \in G.
\end{equation}
Roughly speaking, $\g$ corresponds to fluctuations of $\p$ which takes us to the nearby ground
states with no energy cost, while $\tilde\p$ contains genuine excitations of $\p$. Note that
\cref{tildepdefn} is invariant under $\tilde\p \ra \cD(h)\tilde\p$ with $h\in H$ and hence
determines $\g$ only up to a coset equivalence $\g \sim \g h$. Let us pick a representative from
each coset $\g = \g(\vf)$ parametrized by a field $\vf$ living in the Lie algebra quotient
$\fg/\fh$, which can be identified as the \emph{Goldstone modes} of the broken symmetry. Under a
$g\in G$ transformation,
\begin{equation}\label{vf.transformation}
  \g(\vf) \ra g \g(\vf) h(\vf,g)^{-1}, \quad
  \tilde\p \ra \cD(h(\vf,g))\tilde\p,
\end{equation}
for some $h(\vf,g) \in H$, such that $\p \ra \cD(g)\p$ and \cref{tildepdefn} remains invariant. From
these transformation properties, it is clear that the theory cannot contain a mass term for $\vf$,
rendering it massless. It follows that $\vf$ substantially affects the low energy fluctuations of
the theory and must be taken into account in the superfluid description. A quick comparison can be
made with the Abelian case, where $G=\rmU(1)$ is broken down to $H = \{1\}$, with
$\g(\vf) = \E{-i\vf}$.  Under a $\E{i\L}\in \rmU(1)$ transformation $\vf \ra \vf - \L$, which is
well known in the Abelian superfluid literature.

For notational purposes, let us introduce a set of generators
$\{\mathrm t_\a\} = \{\mathrm t_i,\mathrm t_a\}$ of $G$ such that the subset $\{\mathrm t_i\}$
generates $H$. We orthonormalize these generators by choosing
$\mathrm t_\a \cdot \mathrm t_\b = \Tr{\mathrm t_\a\mathrm t_\b} = \eta_{\a\b}$, where $\eta_{\a\b}$
is a diagonal matrix with entries $\pm 1$. Given an $X = X^\a \mathrm t_\a\in \fg$, under a $g\in G$
transformation $X\ra \mathrm{Ad}_g(X) = (\mathrm{Ad}_g)^{\a}_{\ \b} X^\b \mathrm t_\a = gXg^{-1}$.

While dealing with partially broken symmetries, we are confronted with an obstacle: the quotient
$\fg/\fh$ is not a Lie Algebra and hence $\vf$ does not transform ``nicely'' under the action of
$G$, which poses a difficulty while formulating superfluids. We circumvent this problem by
introducing a pair of projection operators $\rmP,\overline\rmP:\fg\ra \fg$ as,
\begin{align}\label{PbarP}
  \rmP(X) &= \rmP^{\a}_{\ \b} X^\b \mathrm t_\a = \lb(\mathrm{Ad}_\g)^{\a}_{\ i}
            (\mathrm{Ad}_{\g^{-1}})^{i}_{\ \b} \rb X^\b \mathrm t_\a, \nn\\
  \overline\rmP(X) &= \overline\rmP^{\a}_{\ \b} X^\b \mathrm t_\a = \lb(\mathrm{Ad}_\g)^{\a}_{\ a}
            (\mathrm{Ad}_{\g^{-1}})^{a}_{\ \b} \rb X^\b \mathrm t_\a.
\end{align}
They transform covariantly under the action of $G$, i.e. under a $g\in G$ transformation
$\rmP(X),\overline\rmP(X)\ra \mathrm{Ad}_g(\rmP(X)),\mathrm{Ad}_g(\overline\rmP(X))$. Using these we
can re-bundle the information in $\vf$ into
$\tilde\dow_\mu \vf = \overline \rmP(i\dow_\mu\g(\vf)\g(\vf)^{-1} ) \in \fg$ which transforms
``nicely'' in the Adjoint representation of $G$. Introducing the operators $\rmP$, $\overline\rmP$
will also considerably simplify the notation in the following non-Abelian superfluid analysis,
resulting in a pleasant resemblance with the better known Abelian results. As an added benefit, we
can revert back to ordinary fluids at any point by setting $\overline\rmP = 0$, $\rmP = \id_\fg$
(identity in $\fg$).


\lettersection{Superfluid dynamics}We are interested in studying low energy fluctuations of a theory
with a spontaneously broken internal symmetry. As eluded before, any such description must contain
the Goldstone modes $\vf$ as a dynamical field, with dynamics provided by a
$\dim(\fg/\fh)$-component equation,
\begin{equation}\label{vf-EOM}
  K=0 \in \overline\rmP(\fg).
\end{equation}
Here $K$ depends on the details of the microscopic theory. Allowing for an arbitrary dynamical
equation for $\vf$ is a novel feature of our formalism, which in the conventional treatment of
superfluids is taken to be the ``Josephson equation'' by hand (see
e.g. \cite{Bhattacharya:2011tra}). For us however, this will follow as a constraint from the second
law of thermodynamics. A theory invariant under spacetime translations and $G$ transformations must
also contain an associated conserved \emph{energy-momentum tensor} $T^{\mu\nu}$ and a $\fg$-valued
\emph{charge current} $J^\mu$ in its spectrum. To probe these observables we couple the theory to a
slowly varying metric $g_{\mu\nu}$ and a gauge field $A_\mu$. We denote the covariant derivative
associated with the Levi-Civita connection $\G^\l_{\ \mu\nu}$ by $\N_\mu$, while the gauge covariant
derivative associated with $A_\mu$ and $\G^\l_{\ \mu\nu}$ is denoted by $\Df_\mu$. In presence of
these external sources, respective conservation laws take the form,
\begin{equation}\label{EOM.super.na}
  \N_\nu T^{\nu\mu} = F^{\mu\nu} \cdot J_\nu + \xi^\mu \cdot K + \rmT^{\mu\perp}_{\rmH}, \quad
  \Df_\mu J^\mu = \rmJ^\perp_\rmH - K,
\end{equation}
where we have allowed for $\vf$ to go offshell ($K \neq 0$).
$F_{\mu\nu} = 2\dow_{[\mu} A_{\nu]} - i[A_{\mu},A_{\nu}] \in \fg$ is the gauge field strength and
$\xi_\mu = \overline\rmP(A_\mu) + \tilde\dow_\mu \vf \in \overline\rmP(\fg)$ is called the
\emph{superfluid velocity}. The Hall currents $\rmT_\rmH^{\mu\perp}$, $\rmJ_\rmH^\perp$ represent
the contribution from possible gravitational and flavor anomalies in the microscopic theory
respectively.
If the conservation laws (\ref{EOM.super.na}) are unfamiliar to the reader, one way to derive them
is to consider a field theory effective action $S[g_{\mu\nu},A_\mu,\vf]$, and parametrize its
infinitesimal variation as,
\begin{equation}
  \d S = \int \{\df x^\mu\} \sqrt{-g} \Big[ \half T^{\mu\nu} \d g_{\mu\nu} + J^\mu \cdot \d A_\mu
  + K \cdot \tilde\d\vf \Big],
\end{equation}
where $g = \det g_{\mu\nu} $ and $\tilde \d\vf = \overline\rmP\lb i\d\g(\vf) \g(\vf)^{-1} \rb$.
Given this setup, one can check that the conservation laws (\ref{EOM.super.na}) are merely the Ward
identities corresponding to infinitesimal diffeomorphisms and $G$ gauge transformations.

The conservation laws (\ref{EOM.super.na}) can provide dynamics for a theory formulated in terms of
the \emph{hydrodynamic fields}: normalized 4-velocity $u^\mu$ (with $u^\mu u_\mu = -1$), temperature
$T$ and chemical potential $\mu\in\fg$, in addition to the Goldstone modes $\vf$. It should be noted
however that these are merely some fields chosen to describe the system, and like in any field
theory, can admit an arbitrary redefinition; we will return to this issue later. In general, the
observables $T^{\mu\nu}$, $J^\mu$, $K$ appearing in \cref{vf-EOM,EOM.super.na} can have an arbitrary
dependence on the fields $\P = \{u^\mu,T,\mu,g_{\mu\nu},A_\mu,\xi_\mu\}$. In hydrodynamics however,
we are only interested in the low energy fluctuations of the constituent fields $\P$, which can be
translated as the configurations of $\P$ that admit a perturbative expansion in derivatives. This
allows us to write down the most generic allowed expressions for $T^{\mu\nu}$, $J^\mu$, $K$ in terms
of $\P$ truncated up to a finite order in derivatives, called the \emph{superfluid constitutive
  relations}. At a given order, constitutive relations will contain all the possible tensor
structures allowed by symmetry (modulo field redefinitions) called \emph{data}, multiplied with
arbitrary scalars called \emph{transport coefficients}. The explicit functional form of these
transport coefficients depends on the underlying microscopic theory, and can be computed using the
Kubo formula \cite{Kubo:1957mj} in linear response theory. Even without knowledge of the microscopic
theory however, we can put some stringent constraints on the transport coefficients by imposing some
physical requirements such as a local version of the second law of thermodynamics,

\emph{``Given a set of constitutive relations $T^{\mu\nu}$, $J^\mu$, $K$, there must exist an
  entropy current $J^\mu_S$ whose divergence is non-negative, i.e. $\N_\mu J_S^\mu \geq 0$, for all
  the superfluid configurations satisfying the conservation laws (\ref{EOM.super.na}).''}

It is worth pointing out that this statement is slightly stronger than the one used previously in
the superfluid literature (e.g. \cite{Bhattacharya:2011tra}), as it is imposed even when $\vf$ is
offshell. This extra information fixes \cref{vf-EOM} to be the Josephson equation, as we will now
illustrate.

\emph{Ideal superfluids.}---Consider the most generic constitutive relations and
entropy current of a superfluid at zero derivative order,
\begin{align}\label{ideal.sf.onshell}
  T^{\mu\nu} &= (\e + P)u^\m u^\n + P g^{\m\n} + \xi^\m \cdot \r_s \cdot \x^\n, \nn\\
  J^\m &= q u^\m + q_s \cdot \x^\mu, \qquad
  J_S^\m = s u^\m + s_s \cdot \xi^\mu,
\end{align}
along with a scalar $K$. We have fixed the ideal order definition of $u^\mu$ by eliminating a term
like $\e_s\cdot \xi^{(\mu} u^{\nu)}$ from $T^{\mu\nu}$. On the other hand, ideal order definitions
of $T$, $\mu$ are fixed via the \emph{first law of thermodynamics},
\begin{equation}\label{firstlaw}
  \df \e = T \df s + \mu^\a \Df q_\a + \half f_{\a\b} \Df (\xi^{\mu,\a}\xi_\mu^\b),
\end{equation}
where we have defined $f\in \fg \times \fg$ (with $f^{\a\b} = f^{\b\a}$ and
$\rmP^{\g}_{\ \a} f^{\a\b} = 0$).  Using the conservation laws (\ref{EOM.super.na}) and imposing
$\N_\mu J_S^\mu \geq 0$ we can find the following constraints,
\begin{align} 
  \e &= sT + q \cdot \mu - P \quad \text{(\emph{Euler relation})}, \nn\\
  K &= - \frac{\a}{T} \lb u^\mu\xi_\mu - \overline\rmP(\mu)\rb
      + \Df_{\mu} (f \cdot \xi^\mu) + i [\xi_\mu,f\cdot\xi^\mu], \nn\\
  s_s &= 0, \quad \r_s = - q_s = f, \quad \mu \cdot i[\xi_\mu,f\cdot\xi^\mu] = 0,
\end{align}
for some $\a \geq 0$. Plugging these back into \cref{ideal.sf.onshell}, we get the constitutive
relations of an ideal non-Abelian superfluid. The surviving coefficients can be interpreted as:
\emph{pressure} $P$, \emph{energy density} $\e$, \emph{charge density} $q$, \emph{entropy density}
$s$ and \emph{superfluid density} $f$. Setting $K=0$ we recover the non-Abelian Josephson equation
as promised,
\begin{equation}\label{josephson}
  u^\mu\xi_\mu = \overline\rmP(\mu) + \frac{T}{\a} \lb \Df_{\mu} (f \cdot \xi^\mu) + i [\xi_\mu,f\cdot\xi^\mu] \rb + \cO(\dow).
\end{equation}
In the Abelian case, it reduces to its well known form (with few corrections)
$u^\mu\xi_\mu = \mu + \frac{T}{\a} \N_\mu (f \xi^\mu) + \cO(\dow)$. Interestingly, this equation
showed up in the equilibrium analysis of \cite{Bhattacharyya:2012xi} disguised as
$\N_\mu (f \xi^\mu) = 0$, which was unrecognizable as the Josephson equation.


\lettersection{Offshell formalism for superfluids}Having worked out the ideal superfluids, we can in
principle extend this procedure to constitutive relations with arbitrarily high number of
derivatives. However, implementing the second law becomes messier as we go higher in the derivative
expansion, because at a given order in derivatives we are required to use the lower order
conservation laws before imposing $\N_\mu J_S^\mu \geq 0$ (see
e.g. \cite{Banerjee:2015vxa}). Fortunately, as realized by \cite{Loganayagam:2011mu} for ordinary
fluids, it is possible to extend the second law to cases where the conservation laws are not
satisfied (i.e. superfluid is kept in contact with an external bath), by adding arbitrary
combination of the conservation laws (\ref{EOM.super.na}) to $\N_\mu J_S^\mu$,
\begin{multline}\label{offshell2law_superfluid}
  \N_\mu J_S^\mu +
  \b_\mu\lb \N_\nu T^{\nu\mu} - F^{\mu\nu} \cdot J_\nu - \xi^\mu \cdot K - \rmT^{\mu\perp}_{\rmH} \rb \\
  + \nu \cdot \lb \Df_\mu J^\mu + K - \rmJ^\perp_\rmH \rb \geq 0.
\end{multline}
Here $\b^\mu$, $\nu$ are some arbitrary fields. Let
us define $N^\mu = J_S^\mu + \b_\nu T^{\nu\mu} + \nu \cdot J^\mu$ and
$\rmN^\perp_\rmH = \b_\mu \rmT^{\mu\perp}_\rmH + \nu \cdot \rmJ^\perp_\rmH$. In terms of these,
\cref{offshell2law_superfluid} can be recasted in a more useful form,
\begin{equation}\label{free.energy.eqn}
  \N_\mu N^\mu - \rmN_\rmH^\perp - \D = \F \cdot \scrC,
\end{equation}
where $\D$ is a positive definite quadratic form. To make the notation compact we have introduced,
\begin{equation}\label{Phi-definition}
  \scrC = \begin{pmatrix}
    T^{\mu\nu} & J^\r & K
  \end{pmatrix}, \quad \F = \begin{pmatrix} \half \d_\scrB g_{\mu\nu} & \d_\scrB A_\r &
    \tilde\d_\scrB\vf
  \end{pmatrix},
\end{equation}
which are vectors in the composite space
$\fV = \text{(sym. tensor)} \oplus (\fg \times \text{vector}) \oplus \overline\rmP(\fg)$.
``$\d_\scrB$'' denotes an infinitesimal diffeomorphism and $G$ gauge transformation with parameters
$\scrB=\{\b^\mu$, $\L_\b = \nu - A_\mu \b^\mu\}$,
\begin{align}
  \d_\scrB g_{\mu\nu} &= \lie_\b g_{\mu\nu} = 2 \N_{(\mu} \b_{\nu)}, \nn\\
  \d_\scrB A_\mu &= \lie_\b A_\mu + \dow_\mu \L_\b - i [A_\mu,\L_\b]
                   = \Df_\mu \nu + \b^\nu F_{\mu\nu}, \nn\\
  \tilde\d_\scrB\vf &= \overline\rmP\lb i\d_\scrB\g(\vf) \g(\vf)^{-1}\rb \nn\\
                      &= \overline\rmP\lb i\lie_\b\g(\vf) \g(\vf)^{-1} + \L_\b\rb = \b^\mu
                        \xi_{\mu} - \overline\rmP(\nu).\nn
\end{align}
One can check that the ideal order definitions of $u^\mu$, $T$, $\mu$ (given around \cref{firstlaw})
imply the relations $\b^\mu = u^\mu/T$, $\nu = \mu/T$ at ideal order. We fix the remaining ambiguity
in the fluid fields by assuming these relations to hold at all orders in the derivative
expansion. Having done that, the allowed superfluid constitutive relations are the most generic
expressions with $T^{\mu\nu}$, $J^\mu$, $K$ in terms of $\P$ which satisfy \cref{free.energy.eqn}
for some $N^\mu$ and $\D\geq 0$.

Note that it is always possible to write down terms $N^\mu_\rmS \in N^\mu$ whose divergence is
either zero or is balanced by some counter terms $\D_\rmS \in \D$, i.e
$\N_\mu N^\mu_\rmS = \D_\rmS$.  We refer to these terms as Class S. They are not genuine
(super)fluid transport, instead they parametrize the multitude of entropy currents which satisfy the
second law for the same set of constitutive relations.

We split the tensor structures that can appear in the constitutive relations into two sectors:
``non-hydrostatic data'' (independent data that contains at least one instance of ``$\d_\scrB$'')
and ``hydrostatic data'' (largest collection of independent data with no non-hydrostatic linear
combination). The second law, similar to the known results in ordinary fluids
\cite{Bhattacharyya:2013lha,Bhattacharyya:2014bha}, imposes strict equality constraints in the
hydrostatic sector, while in the non-hydrostatic sector it only gives a few inequalities at the
first order in derivatives and none thereafter. We will present a quick proof of this statement; in
the hydrostatic sector we will closely follow \cite{Haehl:2015pja} with appropriate modifications
for superfluids, while in the non-hydrostatic sector our presentation will be independent and
simpler.


\emph{Hydrostatic sector.}---Consider the most generic constitutive relations
$\scrC = \scrC_{\text{hydrostatic}}$ which are solely made up of the hydrostatic data. For these,
every independent term in the RHS of \cref{free.energy.eqn} will contain exactly one bare (isn't
acted upon by a derivative) $\d_\scrB$. Hence the associated $N^\mu$ also must contain the
hydrostatic data only, otherwise $\N_\mu N^\mu$ will either be void of a bare $\d_\scrB$ or will
contain multiple ``$\d_\scrB$''. The most generic $N^\mu$ in the hydrostatic sector can therefore
be written as,
\begin{equation}
  N^\mu_{\text{hydrostatic}} = \lb \cN \b^\mu + \Q^\mu_\cN \rb + \bbN^\mu,
\end{equation}
where $\bbN^\mu u_\mu = 0$. $\cN$ is the most generic scalar made out of the independent hydrostatic
data, modulo the total derivative terms. $\Q^\mu_\cN$ is a $\cN$ dependent non-hydrostatic vector defined via,
\begin{equation}\label{HS}
  \N_\mu (\cN \b^\mu) = \frac{1}{\sqrt {-g}}\d_\scrB \lb\sqrt{-g} \cN\rb
  = \F \cdot \scrC_{\rmH_S} -
  \N_\mu \Q^\mu_\cN,
\end{equation}
which ensures that $\N_\mu (\cN \b^\mu + \Q^\mu_\cN)$ has a bare $\d_\scrB$. \Cref{HS} also defines
the the constitutive relations $\scrC_{\rmH_S}$ associated with $\cN$, called Class
$\rmH_S$. $\bbN^\mu$ on the other hand is the most generic hydrostatic vector transverse to $u^\mu$,
such that $\N_\mu\bbN^\mu - \rmN_\rmH^\perp$ has exactly one bare $\d_\scrB$. This requirement
happens to completely determine $\bbN^\mu$ up to some constants, which includes the terms
responsible for anomalies. The easiest way to find $\bbN^\mu$ is using a (transcendental) anomaly
polynomial \cite{Jensen:2013kka,Jensen:2012kj}, which is written only in terms of the curvature
$R^\mu{}_{\nu\r\s}$, field strength $F_{\mu\nu}$ and an auxiliary $\rmU(1)_\sfT$ field strength
$F_{\mu\nu}^\sfT = 2\dow_{[\mu}A_{\nu]}^\sfT$. It follows that $\bbN^\mu$ is independent of $\vf$
and hence is ignorant of the fluid being in the superfluid phase. It allows us to directly import
$\bbN^\mu$ and the respective Class $\rmH_V \cup \rmA$ constitutive relations
$\scrC_{\rmH_V} + \scrC_{\rmA}$ from the ordinary fluid literature \cite{Haehl:2015pja}, where Class
A is the contribution from anomalies.
$\scrC_{\text{hydrostatic}} = \scrC_{\rmH_S} + \scrC_{\rmH_V} + \scrC_{\rmA}$ are therefore the most
generic hydrostatic constitutive relations compatible with the second law. Comparing these to the
most generic expressions allowed by symmetry, we can read out the equality constraints. It is worth
pointing our that these constraints can also be generated using an equilibrium effective action
\cite{Bhattacharyya:2012xi}.

\emph{Non-hydrostatic sector.}---This sector of hydrodynamics contains constitutive relations
$\scrC = \scrC_{\text{non-hydrostatic}}$ which are purely made of the non-hydrostatic data. Since
every non-hydrostatic data has at least one $\d_\scrB$, it can be written as a differential operator
acting on $\F$ defined in \cref{Phi-definition}. Introducing a symmetric covariant derivative
operator $\Df^n = \Df_{(\mu_{1}}\ldots \Df_{\mu_{n})}$ (anti-symmetric derivatives can be
represented by curvature and field strength), the most generic non-hydrostatic constitutive
relations can therefore be written in a compact form,
\begin{equation}\label{non-hydrostatic}
  \scrC_{\text{non-hydrostatic}}
  = - \sum_{n=0}^\infty \half\big[\fC_n \cdot (\Df^n\F) + \Df^n (\fC_n\cdot\F) \big].
\end{equation}
$\fC_n \in \fV \times \fV$ are matrices with additional $n$ symmetric indices to be contracted with
$\Df^n$. The last term in \cref{non-hydrostatic} is taken purely for convenience and can be absorbed
into the first via differentiation by parts. Let us factor $\scrC_{\text{non-hydrostatic}}$ into a
dissipative (Class D) and a non-dissipative (Class $\overline\rmD$) part parametrized by,
\begin{equation}
  \fD_n = \half\lb \fC_n + (-)^n\fC^\rmT_n \rb, \
  \overline\fD_n = \half\lb \fC_n
  - (-)^n\fC^\rmT_n \rb,
\end{equation}
respectively. The nomenclature can be justified by multiplying \cref{non-hydrostatic} with $\F$
giving us (see also \cite{Bhattacharyya:2014bha,Bhattacharyya:2013lha}),
\begin{equation}\label{DbarD}
  \F\cdot\scrC_{\rmD} = - \D_\rmD + \N_\mu N^\mu_{\rmD}, \qquad
  \F\cdot\scrC_{\overline\rmD} = \N_\mu N^\mu_{\overline\rmD},
\end{equation}
where $N^\mu_{\rmD}$, $N^\mu_{\overline\rmD}$ are some vectors gained via successive differentiation
by parts. $\D_\fD$ however is given as,
\begin{equation}
  \D_{\rmD}
  = (\U \F) \cdot \fD^{(0)}_0 \cdot (\U \F),
\end{equation}
where $\U = \sum_{d=0}^\infty \U_d: \fV \ra \fV$ is a differential operator defined by
($\fD^{(n)}_0$ is the part of $\fD_0$ with $n$ number of derivatives, and ``$\dagger$'' denotes the
conjugate of a differential operator:
$\F_1\cdot (\cO\F_2) = (\cO^\dagger\F_1)\cdot \F_2 + \N_\mu(\cdots)^\mu$),
\begin{align}
  \mathrm \U_{d+1} \Big\vert_{d=1}^\infty
  &= - (\fD_0^{(0)})^{-1} \cdot \lB \sum_{k=1}^{d-1} \mathrm \U_k^\dagger
    + \half  \mathrm \U_d^\dagger  \rB \lb \fD^{(0)}_0 \cdot \mathrm \U_d \rb, \nn\\
  \mathrm \U_0 = 1, \quad&
  \mathrm \U_1
  = \half (\fD_0^{(0)})^{-1} \cdot \sum_{n=1}^\infty \lb \fD^{(n)}_0 + \fD_n \Df^n \rb.
\end{align}
Comparing \cref{DbarD,free.energy.eqn}, we can see that Class $\overline\rmD$ constitutive relations
satisfy the second law with $N^\mu = N^\mu_{\overline\rmD}$ and $\D = 0$, hence the name
non-dissipative. On the other hand, dissipative Class D constitutive relations satisfy the second
law with $N^\mu = N^\mu_{\rmD}$ and $\D = \D_{\rmD}$. The condition $\D\geq 0$ implies that all the
eigenvalues of the zero derivative matrix $\fD^{(0)}_{0} \in \fV \times \fV$ are non-negative. It
follows that the only constraints imposed by the second law in non-hydrostatic sector are some
inequalities in Class D at the first order in derivatives.

At the end of the day, we are only interested in describing the superfluid and not its surroundings,
hence the constitutive relations only differing by combinations of the conservation laws must be
identified. It can be verified that for the constitutive relations satisfying
\cref{free.energy.eqn}, the conservation laws (\ref{EOM.super.na}) are purely non-hydrostatic.
Hence without loss of generality, we can use them to eliminate a vector $u^\mu \d_\scrB g_{\mu\nu}$
and a $\fg$-valued scalar $u^\mu \d_\scrB A_\mu$ from the non-hydrostatic data. The upshot of this
is that we can drop the respective terms from $\scrC_{\rmD}$ and $\scrC_{\overline\rmD}$. Had we
eliminated any other data using the conservation laws, the respective constitutive relations would
be related to the current ones, at most, by a field redefinition.


\lettersection{Classification}In our quest of finding the constraints, we have classified the entire
(super)fluid transport compatible with the second law of thermodynamics into 5 mutually distinct
classes: $\rmA$ (anomalies), $\rmH_{S}$ (hydrostatic scalars), $\rmH_{V}$ (hydrostatic vectors),
$\overline\rmD$ (non-hydrostatic non-dissipative) and $\rmD$ (dissipative), along with a Class
$\rmS$ worth of arbitrariness in the associated entropy current.

To compare with the classification of \cite{Haehl:2014zda}, we decompose
Class S into a part with $\D_\rmS = 0$ (Class C) and remaining (Class $\rmS_\rmD$). In the ordinary
fluid limit, Classes A, C, $\rmH_S$, $\rmH_V$ of \cite{Haehl:2014zda} are same as ours by
definition, while their Class D is Class $\rmD \cup \rmS_\rmD$ for us. A major difference between
the two classifications is that our Class $\overline \rmD$ contains (but is not equal to) their
Classes $\rmB \cup \overline\rmH_S \cup \overline\rmH_V$. For completeness, \cite{Haehl:2014zda}
introduced a ``Class B with $\U$ operators'' which can be shown to be equal to our Class
$\overline\rmD$ (and hence containing their own Classes
$\rmB \cup \overline\rmH_S \cup \overline\rmH_V$), but parametrized very differently. It is evident
therefore, that our classification eliminates some of the redundancies inherent in the
classification of \cite{Haehl:2014zda}.  In the dissipative sector, unlike \cite{Haehl:2014zda} our
parametrization allows us to isolate the ``true dissipation'' from mere entropy current
redundancies.  Additionally, our parametrization in \cref{DbarD} of Classes
$\rmD\cup \overline \rmD$ allows us to easily eliminate constitutive relations related to each other
by combinations of equation of motion (interpreted as ``residual field redefinitions'' in
\cite{Haehl:2014zda}).


\lettersection{Outlook}This completes our analysis of the (non-Abelian) superfluid constitutive
relations compatible with the second law of thermodynamics. The results can also be applied to an
ordinary fluid, seen as a special case of a superfluid where no symmetry is broken. Similar to an
ordinary fluid, we find that the second law gives no constraints in the non-dissipative
non-hydrostatic sector, while it only gives inequalities at the first derivative order in the
dissipative sector. In the hydrostatic sector however, we get equality-type constraints at every
derivative order, which can be worked out using an equilibrium partition function. In addition, the
second law also gives us the Josephson equation which governs motion of the Goldstone modes
corresponding to the broken symmetry.

An added benefit of working in the offshell formalism is that it provides a natural setting to write
down an effective action describing (super)fluids. As a prototype, constitutive relations in Class
$\rmH_S$ and their dynamical equations can be obtained from an effective action (see
\cite{Haehl:2015pja} for related details),
\begin{equation}
  S_{\rmH_S} = \int \{\df x^\mu\} \sqrt{-g} \ \cN.
\end{equation}
For the remaining classes, writing down an effective action needs passing to the Schwinger-Keldysh
formalism \cite{Haehl:2015pja,Haehl:2015uoc}. As a prospective direction, it will be interesting to
write down a complete effective action for superfluids that implements analyticity constraints in
the Schwinger-Keldysh formalism. It will also be interesting to connect with the on-going
explorations of effective actions in the holographic context \cite{Crossley:2015tka,deBoer:2015ija}.

In this note we concentrated on fluids with broken internal symmetries. The procedure can also be
extended to the breaking of spacetime symmetries, interpreted as introducing space-time
boundaries/surfaces in the (super)fluid \cite{Armas:2015ssd}. It will be interesting to see how the
second law constrains the surface transport coefficients in (super)fluids, and if there is a natural
extension of the presented classification to surface transport.

Finally, all of the results presented here can easily be extended to Galilean superfluids using the
null fluid formalism of \cite{Banerjee:2015uta,Banerjee:2015hra,Jain:2015jla}. In a companion paper
\cite{galilean-superfluids}, we will use ``null superfluids'' to work out the constraints on Abelian
Galilean superfluid transport up to first order in the derivative expansion.

\countstop


\section{Acknowledgements}

The author would like to thank Nabamita Banerjee, Jyotirmoy Bhattacharya, Suvankar Dutta and Felix
Haehl for extensive discussions on various points presented in this work. Author also wishes to
acknowledge helpful conversations with Michael Appels, J\'{a}come Armas, Leopoldo Cuspinera and Ruth
Gregory during the course of this project. AJ is financially supported the Durham Doctoral
Scholarship offered by Durham University.

\bibliography{aj-bib}

\end{document}